\documentclass[sigconf]{acmart}

\copyrightyear{2025}
\acmYear{2025}
\setcopyright{rightsretained}
\acmConference[ICMI '25]{Proceedings of the 27th International Conference on Multimodal Interaction}{October 13--17, 2025}{Canberra, ACT, Australia}
\acmBooktitle{Proceedings of the 27th International Conference on Multimodal Interaction (ICMI '25), October 13--17, 2025, Canberra, ACT, Australia}
\acmDOI{10.1145/3716553.3757091}
\acmISBN{979-8-4007-1499-3/2025/10}
\AtBeginDocument{%
  }

\acmSubmissionID{1026}
\usepackage{cleveref}
\usepackage{booktabs}
\usepackage{multirow}
\usepackage{array}
\usepackage{booktabs}
\usepackage{xcolor}
\usepackage{enumitem}
\usepackage{colortbl}
\usepackage{hyperref}

\hypersetup{
  colorlinks=true,
  linkcolor=blue,
  urlcolor=blue,
  citecolor=blue
}

\begin{document}

\title{SocialWise: LLM-Agentic Conversation Therapy for Individuals with Autism Spectrum Disorder to Enhance Communication Skills}

\author{Albert Tang}


\acmConference{}{}{}
\acmBooktitle{} 

\acmArticleType{Review}
\acmCodeLink{https://anonymous.4open.science/r/SocialWise-4224/}
\keywords{SocialWise; LLM; Agents; Conversation Therapy; Autism Spectrum Disorder}

\begin{abstract}
Autism Spectrum Disorder (ASD) affects more than 75 million people worldwide. However, scalable support for practicing everyday conversation is scarce: Low-cost activities such as  story reading yield limited improvement. At the same time, effective role-play therapy demands expensive, in-person sessions with specialists. SocialWise bridges this gap through a browser-based application that pairs LLM conversational agents with a therapeutic retrieval augmented generation (RAG) knowledge base. Users select a scenario (e.g., ordering food, joining a group), interact by text or voice, and receive instant, structured feedback on tone, engagement, and alternative phrasing. The SocialWise prototype, implemented with Streamlit, LangChain, and ChromaDB, runs on any computer with internet access, and demonstrates how recent advances in LLM can provide evidence-based, on-demand communication coaching for individuals with ASD.

\end{abstract}

\maketitle

\section{Introduction}
Autism Spectrum Disorder (ASD) affects \emph{1 in 36} children in the United States and more than 75 million people worldwide \cite{maenner2023}. About 40 \% of autistic adults have never held paid employment \cite{shattuck2012}.  
\smallskip

\noindent\textbf{Limits of current support.} Speech therapy improves turn-taking and topic maintenance \cite{parsons2017}, but these services remain constrained by cost, geography, and provider availability \cite{BrookmanFrazee2012_Parent}. Low-cost activities such as story reading yield only modest gains \cite{FletcherWatson2016}, while augmented-reality tools still require therapist supervision \cite{daSilva2015_STAR}. A scalable alternative must deliver realistic role-play \emph{and} personalized feedback without specialist oversight.

\smallskip
\noindent\textbf{Our solution.} SocialWise is a browser-based application that pairs LLM agents with a retrieval-augmented knowledge base. Users choose a scenario (e.g., ordering food), interact via text or voice (OpenAI Whisper STT + TTS), and receive instant, structured feedback tailored to ASD communication goals.

\smallskip
\noindent\textbf{Demo contributions.}
\begin{itemize}[leftmargin=*]
  \item A multimodal web interface that delivers therapist-style role-play for individuals with ASD anytime, anywhere.
  \item An LLM-driven feedback agent grounded by curated therapeutic documents.
  \item Preliminary user data (N = 34) indicating high perceived helpfulness (4.15 / 5) and 100 \% willingness to recommend.
\end{itemize}
In a live demonstration, visitors will experience a full interaction cycle—scenario selection, live conversation, and feedback—in under three minutes, demonstrating how recent LLM advances can make evidence-informed communication coaching broadly accessible.

\section{Background \& Related Work}

\noindent To position SocialWise, we review prior efforts along four aspects: 
traditional role-play interventions, digital communication aides, LLM 
role-play agents, and educational chatbots.

\smallskip
\noindent\textbf{Role-play interventions.} Classical Social Skills Training (SST) programs combine modelling, role-play, and feedback, and remain the gold standard for pragmatic language gains in ASD \cite{ElKassem2024_SST}. Cost and geography, however, restrict access \cite{BrookmanFrazee2012_Parent}.

\smallskip
\noindent\textbf{Digital aides.} Augmentative-and-Alternative Communication (AAC) tools \cite{mirenda2003} and AR systems such as STAR \cite{daSilva2015_STAR} broaden reach yet still require therapist supervision and provide little adaptive feedback.

\smallskip
\noindent\textbf{LLM role-play agents.} Character-centric frameworks (e.g., RoleLLM \cite{Wang2023_RoleLLM} and CharacterGLM \cite{Zhou2024_CharacterGLM}) can sustain persona-consistent dialogue but do not target therapeutic coaching.

\smallskip
\noindent\textbf{Educational chatbots.} AI tutors improve skill rehearsal in translation \cite{Woollaston2023_TAMMY} and interviewing \cite{Chen2021_AISim}, yet lack grounding in ASD communication goals.

\smallskip
\noindent\textbf{Closing the remaining gaps.} \emph{SocialWise} delivers (1) \textbf{adaptive feedback} by coupling GPT-4o-mini dialogue with a retrieval-grounded analysis engine aligned to evidence-based ASD therapy guidelines; (2) seamless \textbf{speech \& text interaction} via Whisper STT and OpenAI TTS, enabling fully multimodal practice; and (3) true \textbf{therapist-free accessibility}, as the entire system runs in a standard web browser without specialised hardware or professional supervision.

\section{System Snapshot}

\noindent\textbf{Architecture.} SocialWise is a Streamlit web app that calls GPT-4o-mini for
dialogue, Whisper STT, and OpenAI TTS.  A LangChain Retrieval-Augmented Generation
pipeline (ChromaDB dense retriever) grounds feedback in ASD-therapy
documents.

\smallskip
\noindent\textbf{Runtime.} Average latency is 1.6s for text and 4.8s for speech on a 2021
M1 laptop; Our system is scalable because of the use of OpenAI’s API and Streamlit’s deployment system, and we will use FastAPI with NextJS for future production.

\smallskip
\noindent\textbf{Safety.} Hallucination is reduced by RAG grounding, instruction prompts,
and expert review; CoT bias-mitigation prompts are enabled, with fine-tuning planned.


\begin{figure*}[t]
  \centering
  \includegraphics[width=0.9\linewidth]{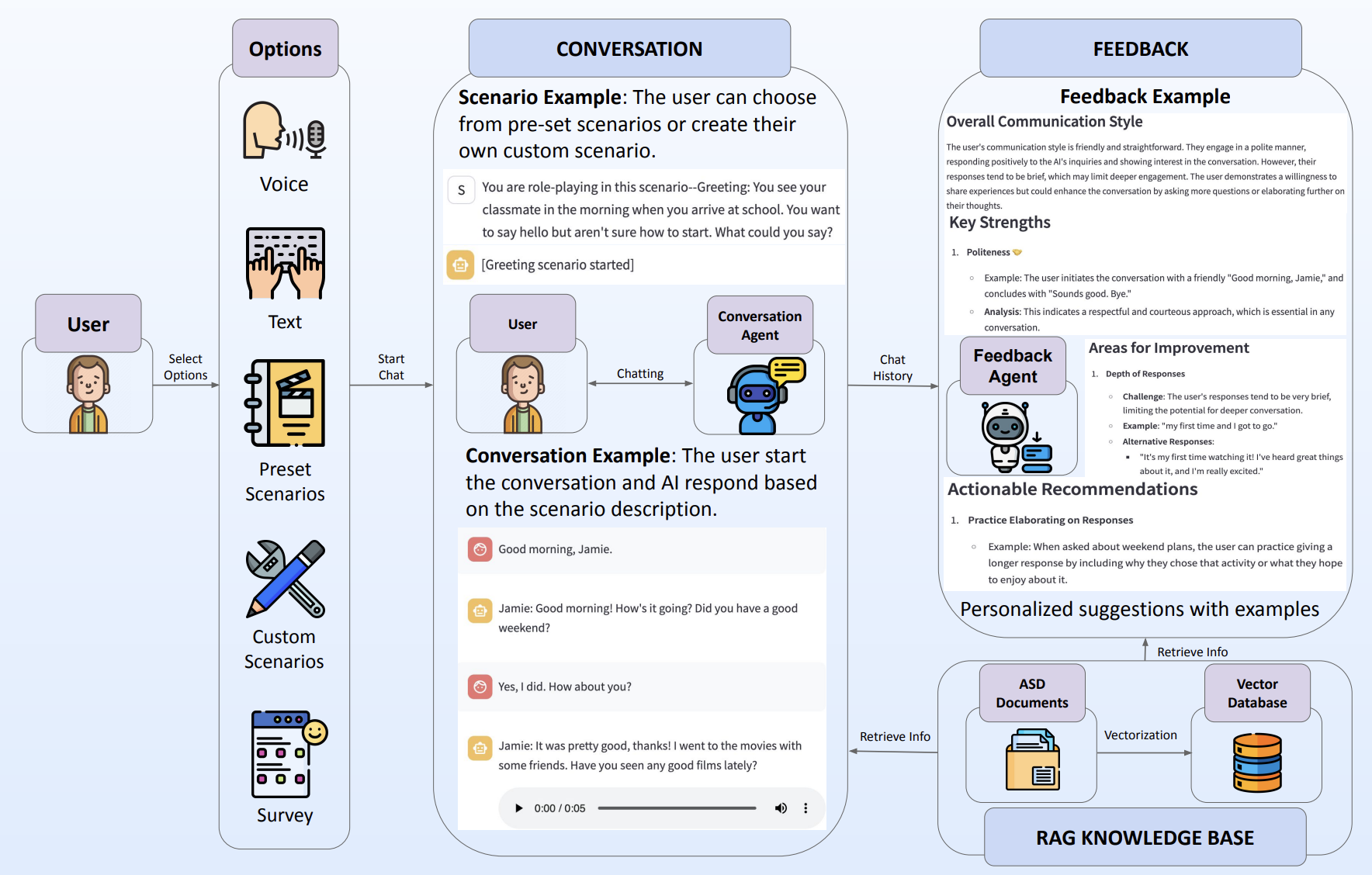}
  \caption{The workflow of SocialWise, an AI-powered conversation-therapy chatbot that helps individuals with autism practise social interactions. The user selects conversation options, role-plays with a conversation agent, and the system retrieves relevant knowledge from a RAG-powered vector database. A feedback agent analyses user responses and provides constructive feedback.}
  \Description{Block-diagram of the SocialWise pipeline. From left to right: (1) a user interface where the learner chooses a scenario; arrow to (2) an AI conversation agent that exchanges dialogue with the learner; arrow down to (3) a feedback agent that scores and explains the learner’s communication performance; final arrow to (4) a retrieval-augmented generation (RAG) vector-database module labelled “Knowledge Retrieval”; arrow back up to the conversation agent and the feedback agent. All blocks are connected with arrows indicating the data flow.}
  \label{fig:pipeline}
\end{figure*}

\section{Demo Interaction Flow}

\begin{figure*}[ht]
  \centering
  \includegraphics[width=0.9\linewidth]{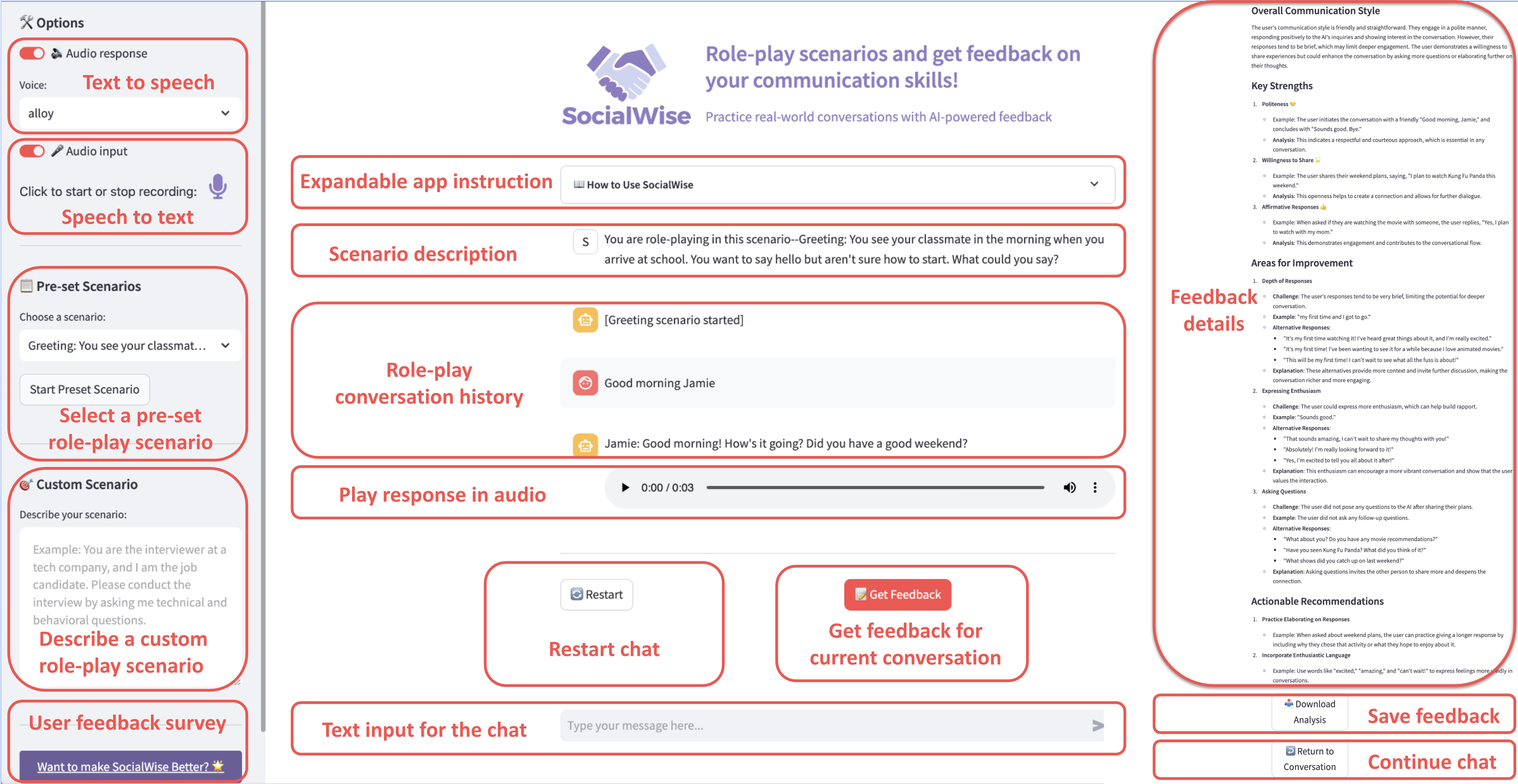}
  \caption{Annotated screenshot of the SocialWise web interface.}
  \Description{Screenshot of the SocialWise web app with red call-out boxes labelling key UI regions. A narrow left sidebar titled “Options” offers toggles for text-to-speech and speech-to-text, a dropdown to pick a pre-set scenario with a “Start Preset Scenario” button, and a text area for defining a custom scenario. The wide center column begins with the SocialWise logo and tagline, then an expandable “How to Use SocialWise” instruction bar, a scenario description panel, the role-play chat history, an audio playback bar, “Restart” and “Get Feedback” buttons, and a text‐input field. The right-hand panel shows detailed AI feedback—overall communication style, key strengths, areas for improvement, and actionable recommendations. Bottom-right buttons let the user download the analysis, save feedback, or continue the chat.}
  \label{fig:UI}
\end{figure*}

\noindent Figure~\ref{fig:UI} shows that a user's journey involves four major steps:
\begin{enumerate}[leftmargin=*,
                itemsep=2pt]   
  \item \emph{Scenario pick.} Visitors select a preset role-play scenario (e.g.\ \textit{“Ordering food”}) or type their own situation.
  \item \emph{Live role-play.} They speak or type; GPT-4o-mini answers in natural speech and on-screen text while retaining context.
  \item \emph{Instant feedback.} Pressing \textsc{Get Feedback} triggers retrieval-grounded analysis that appears in under three seconds, highlighting strengths, areas to improve, and alternative phrasings.
  \item \emph{Iterate / restart.} Attendees can continue the conversation, switch scenarios, or export the feedback as HTML.
\end{enumerate}

\noindent Figure~\ref{fig:UI} shows an annotated screenshot of the SocialWise web interface.
The left sidebar (\textit{Options}) lets users toggle text-to-speech and speech-to-text, pick a pre-set
scenario, or type a custom scenario; a link at the bottom opens the user-feedback survey.
The centre column displays (top-to-bottom) an expandable “How to Use” panel, the scenario description,
the live role-play history, an optional audio-playback bar, buttons to restart the chat or request
feedback, and a text-input field.
The right column presents the structured feedback: overall style, key strengths, areas for improvement,
and actionable recommendations, with buttons below to download the analysis or return to the
conversation. The feedback is on a separate page on the web app.

The layout makes all critical functions—scenario selection, multimodal conversation, and
evidence-based feedback for the live demo.


\section{Preliminary User Feedback}

We ran a remote pilot with \textbf{34 autistic adults} (ages 18–34).  
After completing at least one role-play, participants rated overall helpfulness \textbf{4.15 / 5} and \textbf{100 \%} said they would recommend the tool to a peer.  
The three most cited benefits were improved \emph{turn-taking (67.6 \%)}, \emph{problem-solving in social conflicts (52.9 \%)}, and \emph{reduced anxiety during real conversations (50.0 \%)}.  
Open-ended suggestions focused on adding gamified rewards and advanced scenarios, confirming user appetite for deeper practice.  
Although not a clinical trial, these early results indicate strong perceived value and engagement.



\section{Future Work}
Future work for SocialWise includes conducting formal clinical validation studies, expanding the scenario library for broader social contexts, developing a mobile app, and implementing user progress tracking features.

\section{Responsible Innovation \& Ethics}

\smallskip
\noindent SocialWise is designed for safe, private, and equitable use.  

\smallskip
\noindent\textbf{Privacy.} All conversation logs remain in the browser’s memory and are flushed on page refresh; no user identifiers or audio recordings are stored server-side.  

\smallskip
\noindent\textbf{Bias mitigation.} We employ chain-of-thought (CoT) bias-reduction prompts at every generation step and plan a follow-up fine-tuning model that addresses community feedback.   

\smallskip
\noindent\textbf{Hallucination control.} Retrieval grounding, strict instruction prompts, and human spot-checks limit factual drift.  

\smallskip
\noindent
We are committed to the responsible development and deployment of SocialWise. This includes ongoing monitoring for potential negative impacts and a commitment to transparency with our users about the capabilities and limitations of the AI.

\bibliographystyle{acm}
\bibliography{SocialWise_References.bib}
\end{document}